\documentclass[twocolumn,preprintnumbers,amsmath,amssymb,floatfix,prd]{revtex4}
\usepackage{epsfig}
\usepackage{dcolumn}
\begin{document}
\title{Probing the Sea Quark Content of the Proton with One-Particle-Inclusive Processes}
\author{Ignacio Borsa}
\email{borsa.ignacio@gmail.com}
\affiliation{Departamento de F\'{\i}sica and IFIBA,  Facultad de Ciencias Exactas y Naturales, 
Universidad de Buenos Aires, Ciudad Universitaria, Pabell\'on\ 1 (1428) Buenos Aires, Argentina}
\author{Rodolfo Sassot}
\email{sassot@df.uba.ar} 
\affiliation{Departamento de F\'{\i}sica and IFIBA,  Facultad de Ciencias Exactas y Naturales, 
Universidad de Buenos Aires, Ciudad Universitaria, Pabell\'on\ 1 (1428) Buenos Aires, Argentina}
\author{Marco Stratmann}
\email{marco.stratmann@uni-tuebingen.de}
\affiliation{Institute for Theoretical Physics, University of T\"ubingen, 
Auf der Morgenstelle 14, 72076 T\"ubingen, Germany}
\begin{abstract}
We investigate the feasibility of constraining parton distribution functions in the
proton through a comparison with data on semi-inclusive deep-inelastic lepton-nucleon
scattering. 
Specifically, we reweight replicas of these distributions according to how
well they reproduce recent, very precise charged kaon multiplicity measurements
and analyze how this procedure optimizes the determination of the sea quark densities
and improves their uncertainties. The results can help to shed new light on 
the long standing question on the size of the flavor and charge symmetry breaking 
among quarks of radiative origin.
An iterative method is proposed and adopted to account for the inevitable 
correlation with what is assumed about the parton-to-hadron fragmentation functions
in the reweighting procedure. It is shown how the fragmentation functions can be
optimized simultaneously in each step of the iteration. As a first case study, we
implement this method to analyze kaon production data.
\end{abstract}
%
%\pacs{13.87.Fh, 13.85.Ni, 12.38.Bx}
%
\maketitle

%%%%%%%%%%%%%%%%%%%%%%%%%%%%%%%%%%%%%
\section{Introduction and Motivation}
%%%%%%%%%%%%%%%%%%%%%%%%%%%%%%%%%%%%%
Parton distribution functions (PDFs) \cite{Butterworth:2015oua}
are an essential ingredient in the 
description of nucleons as needed in the quantitative interpretation
of hadronic hard scattering processes in the framework of 
perturbative Quantum Chromodynamics (pQCD). 
Based on the notion of factorization and
together with the hadronization ``probabilities'' of produced final-state partons 
into observed hadrons, commonly known as fragmentation functions (FFs) \cite{Metz:2016swz}, 
PDFs encode all the non-perturbative details about the hadronic 
structure and the path into confinement relevant in theoretical calculations
of hard interactions.

The growing need to validate increasingly precise non-perturbative approaches aimed at 
describing the partonic structure of nucleons \cite{ref:lattice,ref:models}, 
as well as the requirement of having available
unprecedentedly accurate predictions for a multitude of hadronic processes 
where the Standard Model and possible extensions thereof can be challenged, 
place the pursuit of acquiring a precise knowledge on PDFs 
right the center of the most topical discussions at present \cite{Butterworth:2015oua}. 

In spite of the strenuous efforts and the remarkable degree of sophistication achieved
in the extraction of PDFs  and the quantification of their uncertainties
in the last two decades \cite{ref:pdfsets}, the PDFs of largely radiative origin in the nucleon, 
namely the quark-antiquark pairs produced by gluon splitting, still lack the desired precision.
The results for sets of PDFs obtained in global fits to data by the currently most active collaborations 
may differ, for instance, by up to forty percent for strange quarks
and by up to ten percent for the lighter sea quarks \cite{Butterworth:2015oua}. 

In addition, experimental results with a particular, albeit indirect 
sensitivity to sea quark densities of a certain flavor, 
such as measurements of the leptonic decays of W bosons
at the CERN-LHC \cite{Aad:2012sb,Chatrchyan:2013mza,Aaboud:2016btc}, electroweak 
charged-current deep-inelastic scattering (DIS) experiments 
\cite{Bazarko:1994tt,Goncharov:2001qe,Mason:2007zz,Samoylov:2013xoa}, and
kaon multiplicities in semi-inclusive deep-inelastic scattering (SIDIS) \cite{hermesmult}
seem to suggest divergent answers for the shape and normalization of the strange quark PDF.
More specifically, while (SI)DIS experiments favor a sizable
suppression of strange quarks relative to non-strange antiquarks in the proton of 
the order of 50\%, ATLAS results favor scenarios with almost no or very little 
suppression, while CMS prefers a suppressed strange sea.  

In the most naive approximation, one would expect a charge and SU(3) flavor symmetric 
population of quarks of radiative origin, since they are mainly produced by gluon 
splitting into quark and antiquark pairs, which is flavor blind. 
However, different degrees of breaking of these symmetries have been found in various 
experiments, albeit within very large uncertainties. Several competing mechanisms 
have been proposed in the literature \cite{ref:breaking} 
to explain or motivate such findings but none of them seems to be conclusive.

The main difficulties inherent to fully disentangling the share of different flavors 
in the proton were noticed already in the very early days of the parton model 
\cite{Feynman:1973xc}, as well as a plausible solution: 
utilizing the rich information coming from the SIDIS production 
of hadrons with a different flavor content. 
For instance, charged kaon multiplicities obtained off proton targets at medium to large parton 
momentum fractions $x$ contain more positively charged kaons than negative ones according 
to the relative abundances of $u$ and $\overline{u}$ quarks in the proton, and
also critically depend on their ratio to strange quarks. 
The recent access to precisely measured 
multiplicities in SIDIS for positively and negatively charged pions and kaons, produced 
alternatively off proton and deuteron targets, and within different regions of  
momentum fraction $x$ \cite{hermesmult,Adolph:2016bwc}, allows, in principle, to study the 
flavor dependence of PDFs and FFs at an unprecedented level of accuracy. 
The main caveat of such an approach is that it requires a {\em simultaneous} extraction 
of PDFs and FFs, with the cumbersome requirement consisting of the $\chi^2$ minimization 
of a very large number of correlated parameters in a global QCD fit.
As a first step in that direction, we explore in what follows
a sequential, iterative approach, where one-particle-inclusive measurements are used first 
to refine existing sets of PDFs that are in turn adopted as the ingredient 
in a determination of FFs, subsequently utilized in obtaining a next generation of PDFs. 
In the present paper, this procedure is implemented and adopted for kaon production data.
As we shall discuss in detail, the method is found to converge 
very fast to new optimum sets of PDFs and FFs that provide a more accurate description 
of the data along with a noticeable reduction in the uncertainties.

A key ingredient in the above mentioned approach is the so-called reweighting technique
for PDFs as developed and heavily used by the NNPDF collaboration \cite{Ball:2010gb,Ball:2011gg}. 
This strategy allows one to incorporate retroactively but consistently the information contained in data 
sets that are not included in the original global extraction of PDFs and in the determinations of their
uncertainties.
In particular, this framework avoids a full and time-consuming re-fit
but preserves the statistical rigor of 
the original extraction. The method has already been demonstrated 
in different applications, see, e.g., 
\cite{Ball:2010gb,Ball:2011gg,Armesto:2013kqa,Paukkunen:2014zia}. 

The remainder of the paper is organized as follows: in the next section we briefly 
recall the status of the pQCD description of kaon production in SIDIS, 
the recent experimental findings, how they constrain kaon FFs, 
and how much room is left for refining the PDF set adopted in the calculation.
Next, we sketch the main features of the PDFs reweighting technique 
and how it applies to the case under consideration. 
In Section~III, we discuss in some detail the outcome of our reweighting exercise 
for kaon production and the iterative strategy outlined above and
compare the effects of adopting different sets FFs as the starting point
of our analyses. We briefly summarize the main results in Section~IV.

%%%%%%%%%%%%%%%%
\section{Role of SIDIS data in global fits}
%%%%%%%%%%%%%%%%
%
%%%%%%%%%%%%%%%%%%%%%%%%%%%
\subsection{Status of FFs extractions using SIDIS data}
%%%%%%%%%%%%%%%%%%%%%%%%%%%
%
Recently, the {\sc Compass} experiment at CERN \cite{Adolph:2016bwc} published 
extremely precise SIDIS multiplicity measurements for charged kaons produced off deuteron 
targets. The data are presented with a rather fine binning in the relevant kinematical 
variables. 
Together with the multiplicity results from the {\sc Hermes} experiment at DESY 
\cite{hermesmult} taken off both proton and deuteron targets 
at smaller transferred momentum $Q$ than the COMPASS results though,  
the new data allowed for a very precise determination of charged kaon FFs 
and, for the first time, also their uncertainties in a 
global QCD analysis \cite{dss17}, that supersedes a previous, less sophisticated 
extraction performed almost ten years ago \cite{dss}. 
The global fits \cite{dss17,dss} incorporate also additional information from other non-SIDIS
experiments, semi-inclusive electron-position annihilation (SIA) 
and inclusive hadron production at high transverse momentum $p_T$ in
hadron-hadron collisions, which constrained mainly the total flavor singlet
FF and the gluon FF, respectively \cite{dss17,dss}. 
The specific and unique role of SIDIS data in determining 
flavor and charge separated FFs was demonstrated in \cite{dss17}.

Kaon multiplicities $M_{lp(ld)}^{K^{\pm}}$ 
in lepton-proton ($lp$) or lepton-deuteron ($ld$) scattering
are defined as the ratio of the 
inclusive charged kaon yield in SIDIS and the total DIS cross section 
in the same kinematic bins of parton momentum fraction $x$ and photon virtuality $Q^2$:
\begin{equation}
\label{eq:mult}
M_{lp(ld)}^{K^{\pm}} \equiv \frac{ d\sigma^{K^{\pm}}_{lp(ld)}/dx\,dQ^2\,dz}
{d\sigma_{lp(ld)}/dx\,dQ^2}\;.
\end{equation}
In the global fit \cite{dss17} the two-dimensional projections of the three-dimensional 
multiplicity data \cite{hermesmult} onto both the $z-Q^2$ and 
$z-x$ dependence are considered, for four different bins of the kaon's 
momentum fraction $z$ in the case of the {\sc Hermes} data. {\sc Compass} results are presented 
as a function of $z$ in 9 bins of $x$, each subdivided into various bins in the inelasticity
parameter $y$ that effectively select different $Q^2$-ranges. In total 144 (309) data points for both 
$K^+$ and $K^-$ multiplicities are presented in the case of {\sc Hermes} ({\sc Compass}).

Non-SIDIS data comprise $K^-/K^+$ and $K/\pi$ production ratios measured in proton-proton
collisions by the {\sc Star} collaboration at BNL-RHIC \cite{ref:starratio11} 
and the {\sc Alice} collaboration at the LHC \cite{ref:alicedata} respectively, 
and SIA data from {\sc Belle} \cite{ref:belledata}, {\sc BaBar} \cite{ref:babardata}
as well as some older LEP and SLAC experiments
\cite{ref:alephdata,ref:delphidata,ref:opaldata,ref:slddata,ref:opaleta}. 
The combined use of data on different processes, such as SIA, SIDIS and proton-proton 
collisions, is essential to disentangle much more precisely the flavor, charge, and kinematical 
dependence of the hadronization probabilities. The so obtained FFs inherit
ambiguities according to the experimental uncertainties and the precision of theory 
estimates for the different processes due to the choice of scales, PDFs, etc.
These sources of uncertainties are duly included in the error estimates 
obtained in Ref.~\cite{dss17}.

%
%%%%%%FIGURE 1
%
\begin{figure}[th!]
\vspace*{-0.4cm}
\begin{center}
\hspace*{-0.4cm}
\epsfig{figure=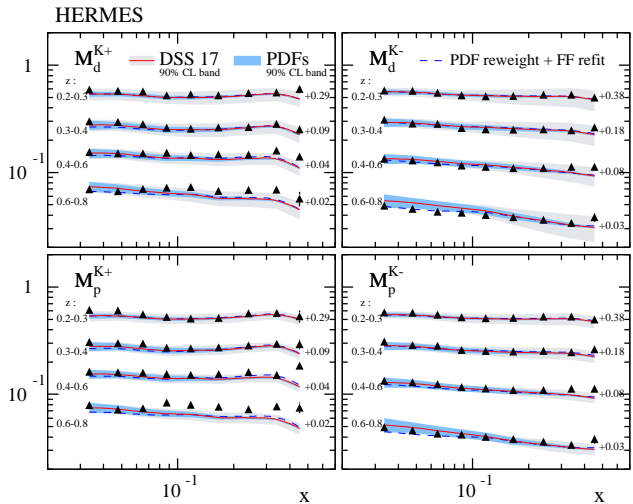,width=0.54\textwidth}
\end{center}
\vspace*{-0.5cm}
\caption{NLO results for charged kaon multiplicities in SIDIS 
computed with the DSS~17 set of FFs \cite{dss17} and the MMHT~14 set of PDFs \cite{Harland-Lang:2014zoa} (solid lines)  
for both proton and deuteron targets compared to data 
from the {\sc{Hermes}} collaboration \cite{hermesmult}.
The light shaded (grey) bands correspond to the uncertainty estimates for the FFs, and
the dark shaded (blue) bands
represent the uncertainties of the MMHT~14 set of PDFs, both at the 
$90\%$ confidence level (C.L.). The dashed lines are the NLO results obtained with a
reweighted set of PDFs and a refined set of FFs, see text.
\label{fig:sidis-hermesx}}
\end{figure} 
The overall agreement between the data and the theoretical calculations based on the DSS~17 set of
FFs \cite{dss17} at next-to-leading order (NLO) accuracy, quantified 
in terms of the usual $\chi^2$ per degree of freedom function, is remarkably good and 
significantly superior than the one achieved with the now outdated DSS~07 set of FFs \cite{dss}.
A closer inspection reveals, however, that the theoretical description is
far from being perfect in the case of quite some of the SIDIS data points, and it cannot
be improved further by allowing for either a greater flexibility in the functional 
parameter space chosen for the input FFs or by relaxing the standard assumptions about the
charge or flavor symmetry of FFs.
The overall $\chi^2$ of the SIDIS data in the DSS~17 analysis
is nevertheless very much acceptable because of the 
rather large PDFs uncertainties in the relevant kinematic regime probed by
{\sc Hermes} and {\sc Compass} which are included as a theoretical uncertainty.
Obviously, this raises the question whether the agreement with SIDIS data
could be improved by fine-tuning the PDFs within their estimated,
rather large uncertainties.

In Fig.~\ref{fig:sidis-hermesx} data on charged kaon multiplicities in SIDIS 
from the {\sc{Hermes}} collaboration \cite{hermesmult}, taken on both proton and deuteron targets,  
are compared to NLO results computed with the DSS~17 set of FFs \cite{dss17}
and the MMHT~14 set of PDFs \cite{Harland-Lang:2014zoa}. The shaded bands 
illustrate the respective uncertainty estimates for the FFs and PDFs.
Roughly speaking, the uncertainty stemming from the chosen set of PDFs is larger for
$K^-$ than for $K^+$ multiplicities mainly because of the ${\overline u}s$ 
content of the $K^-$ meson that predominantly couples to corresponding $\bar{u}$ and $s$
sea quark PDFs compared to the $u{\overline s}$ content of the $K^+$ meson.
The $u$ quark PDF relevant for $K^+$ production is significantly better constrained 
than the ${\overline u}$ sea quark PDF.
Because of this larger dependence on sea quark
PDFs, $K^-$ multiplicities are also typically smaller and fall off faster with 
increasing parton momentum fractions $x$ than their $K^+$ counterparts.
In both cases the PDFs uncertainties grow towards smaller $x$ because of
the relative enhancement of the sea quark content. 

The larger the PDF uncertainties or the differences between data and
theory estimates in Fig.~\ref{fig:sidis-hermesx}, the greater is the potential 
to further constrain them with the SIDIS data. 
Indeed, anticipating the results of our iterative reweighting technique to be discussed
in Sec.~III below, the dashed lines in Fig.~\ref{fig:sidis-hermesx} correspond
to NLO calculations utilizing reweighted sets of PDFs and a 
refined set of FFs that was obtained by using the so improved PDFs in the global analysis.

%
%%%%%FIGURES 2 AND 3
%
\begin{figure*}[th!]
\vspace*{-0.4cm}
\begin{center}
\hspace*{-0.4cm}
\epsfig{figure=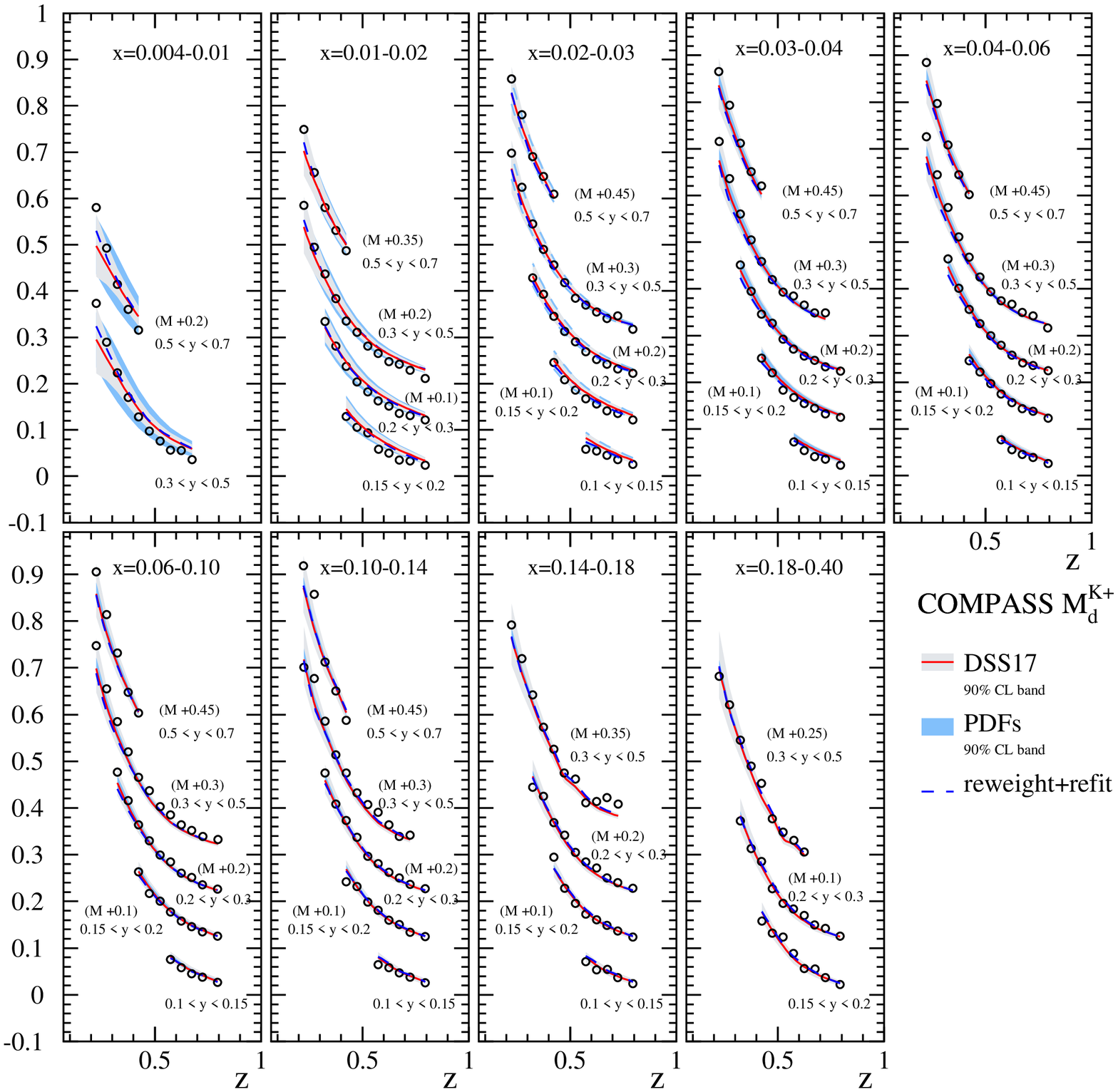,width=0.67\textwidth}
\end{center}
\vspace*{-0.5cm}
\caption{Same as in Fig.~\ref{fig:sidis-hermesx} but now for the
positively charged kaon multiplicities on a deuteron target obtained by 
{\sc{Compass}} \cite{Adolph:2016bwc}; see text.
\label{fig:sidis-compassx}}
%\end{figure*} 
%
%\begin{figure*}[th!]
\vspace*{-0.4cm}
\begin{center}
\hspace*{-0.4cm}
\epsfig{figure=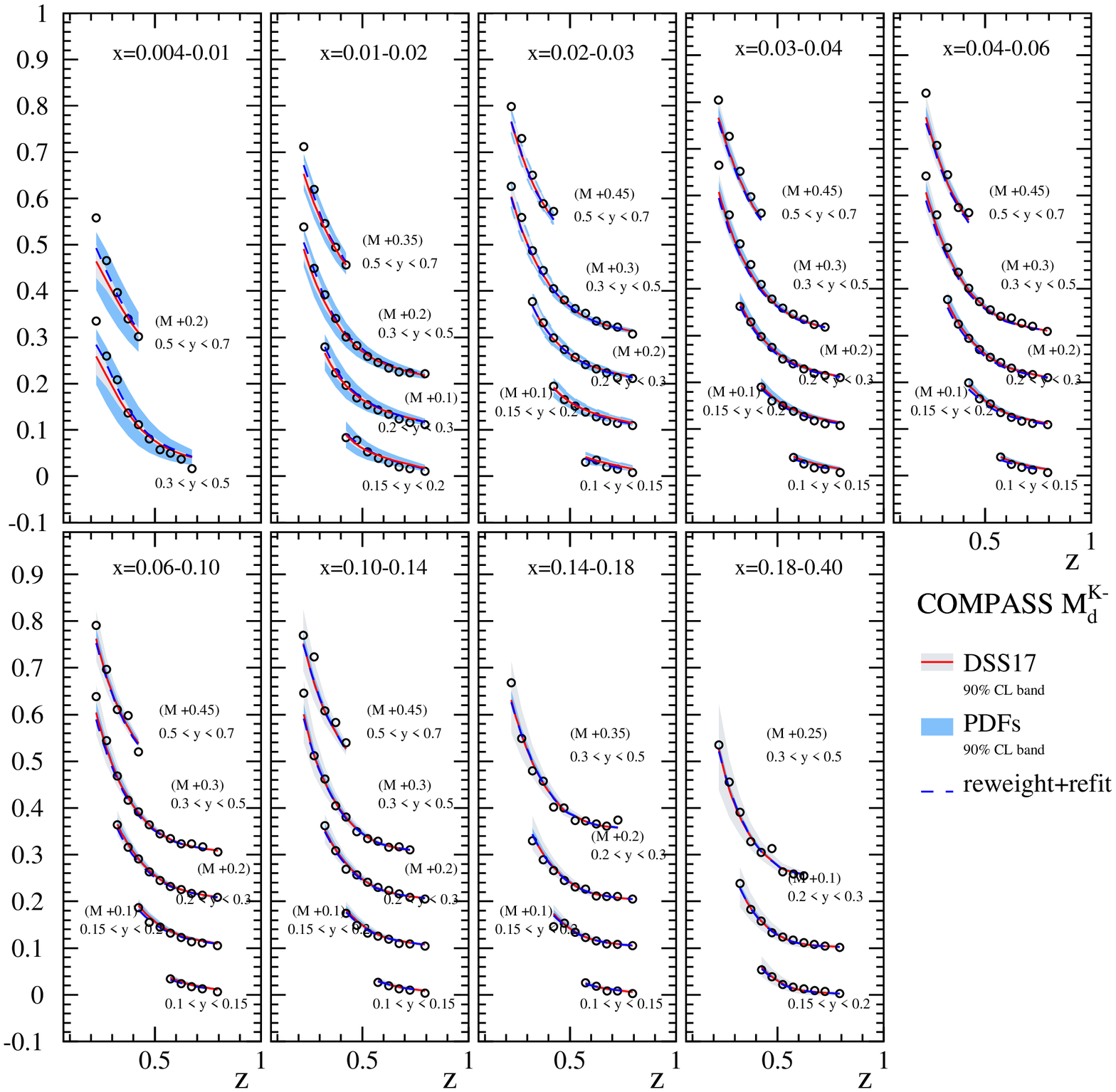,width=0.67\textwidth}
\end{center}
\vspace*{-0.5cm}
\caption{Same as in Fig.~\ref{fig:sidis-compassx} but for negatively 
charged kaons.
\label{fig:sidis-compassxmin}}
\end{figure*} 

Figures~\ref{fig:sidis-compassx} and \ref{fig:sidis-compassxmin} show
the corresponding SIDIS multiplicities for charged kaons off a deuteron 
target as a function of the hadron momentum fraction $z$ as measured 
by {\sc COMPASS}.
Each panel corresponds to a different bin in $x$
and, consequently, probes the PDFs in a different kinematic region. 
Specifically, the first three panels correspond to $x$ values lower than those
covered by the {\sc{Hermes}} data shown in Fig.~\ref{fig:sidis-hermesx}. 
Within each $x$-bin, the data are divided in bins of the inelasticity
parameter $y$ that corresponds to different photon virtualities $Q^2$. Again, 
the data extend the range on $Q^2$ as covered by {\sc{Hermes}}.

%%%%%%%%%%%%%%%%%%%%%%%%%%%
\subsection{PDFs reweighting for SIDIS data}
%%%%%%%%%%%%%%%%%%%%%%%%%%%
%
The reweighting technique allows one to incorporate consistently the 
information contained in new data into an existing set of PDFs without 
the need of refitting them but preserving the statistical rigor of 
its extraction \cite{Ball:2010gb,Ball:2011gg}. 
The method has already been successfully demonstrated in different 
applications. Here, we briefly recall its main features which are
needed for our discussions below.

The method is based on statistical inference and starts
with the generation of a large ensemble of PDF replicas $f_k$, each fitted
to a data replica generated according the experimental uncertainties and 
correlations. The ensemble forms an accurate representation of the
probability distribution of PDFs, and any quantity ${\cal O}$ depending
on the PDFs can be evaluated by averaging the results for the individual 
replicas:
\begin{equation}
\label{eq:mean}
\langle{\cal O} \rangle = \frac{1}{N} \sum_{k=1}^{N}{\cal O}[f_k]\,,
\end{equation}
with $N$ the number of replicas. Using Bayesian inference it is possible
to update the original probability distribution to a new one that accounts
for the information contained in a new measurement, by assigning a new weight
$w_k\neq 1$ for each replica, which measures its agreement with the new data.
The updated estimate for any quantity then becomes
\begin{equation}
\label{eq:newmean}
\langle {\cal O} \rangle_{\text{new}} = \frac{1}{N} \sum_{k=1}^{N} w_k\,{\cal O}[f_k]\,.
\end{equation}

The Bayesian reweighting is fully equivalent to a full new fit provided the
new data set is not too constraining, so that the effective number of replicas,
i.e.\ those with a non-negligible weight $w_k$, is still large enough. 
Two consistency checks are usually performed in order 
to verify that a new set of data can be satisfactorily described by a large enough 
number of replicas in the initial ensemble. The first one is the direct examination 
of the $\chi^2$ profile for the new set of data 
normalized by the number of data $N_{\text{data}}$ before and after the reweighting.
Alternatively, it is possible to estimate by how much the agreement between data 
and theory improves by rescaling the uncertainties of the new data by a factor 
$\alpha$, whose probability is proportional to
\begin{equation}
\label{eq:p_alpha}
{\cal P} (\alpha)  \propto \frac{1}{N} \sum_{k=1}^{N} w_k(\alpha) \,{\cal O}[f_k]\,.
\end{equation}
If ${\cal P} (\alpha)$ is peaked near one, the new data is consistent with
the initial theory distribution, while if it is peaked at a larger value, it
may suggest that uncertainties have been underestimated or the underlying 
theory is not adequate to describe the newly included measurement. 

The NNPDF collaboration provides an ensemble of 1000 replicas for its PDF
set NNPDF~3.0 \cite{Ball:2014uwa} which we utilize in the following to compute weights $w_k$
according to how well each replica $k$ reproduces the {\sc{Compass}} and {\sc{Hermes}} data
for charged kaon production.
The SIDIS multiplicities are computed at NLO accuracy by convoluting 
the recent DSS~17 set of parton-to-kaon FFs \cite{dss17} with each of the NNPDF~3.0 replicas 
and the appropriate NLO coefficient functions as was described, for instance, 
in Refs.~\cite{dss,dss17}.

%
%%%%FIGURE 4
%
\begin{figure}[h!]
\vspace*{-0.2cm}
\begin{center}
\hspace*{-0.4cm}
\epsfig{figure=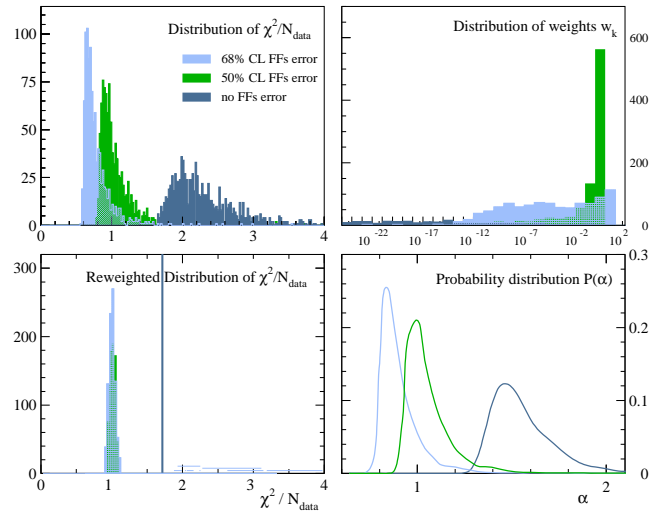,width=0.52\textwidth}
\end{center}
\vspace*{-0.5cm}
\caption{Distribution of $\chi^2$ per data point for the combined kaon multiplicity 
data sets of {\sc Hermes} and {\sc Compass} before and after reweighting,
upper and lower left panel, respectively,
the weights $w_k$ (upper right panel) and of ${\cal P} (\alpha)$ (lower right panel). 
Each distribution is computed neglecting the error coming from the FFs (green),
including the 68\% C.L.\ estimates of Ref.~\cite{dss17} (blue), 
and an intermediate C.L.\ value adjusted such that ${\cal P} (\alpha)$ 
peaks at one (red); see text.
\label{fig:histo}}
\end{figure} 
The agreement between the theoretical estimate and data not only depends
on the quoted experimental errors and on the uncertainties of the chosen set of PDFs 
but also on the ambiguities inherent to the set of parton-to-kaon FFs used in the computation.
Therefore, it seems reasonable to include also the uncertainty estimates on the
FFs given in Ref.~\cite{dss17} in the $\chi^2$ estimate by propagating them to 
the SIDIS multiplicities. 
However, this procedure necessarily introduces some ambiguity: 
the uncertainty estimate of FFs as provided by Ref.~\cite{dss17}
already accounts for the errors of the SIDIS data and that
of the particular PDF set employed in their extraction.
This implies a certain degree of double counting of uncertainties. 
On the other hand, excluding the error on FFs altogether would disregard 
legitimate sources of uncertainties in the reweighting procedure, 
such as those coming from other, non-SIDIS data sets used in the determination
of FFs. Of course, this conceptual problem would not be present in a simultaneous global
analysis of PDFs and FFs. 

In Fig.~\ref{fig:histo} we show the $\chi^2/N_{\text{data}}$ distribution of the PDF 
replicas before and after their reweighting (upper and lower left panels, respectively), 
the scale factor ${\cal P}(\alpha)$ probability (lower right panel) and the 
distribution of weights $w_k$ (upper right panel), for three different options 
regarding the FFs contribution to the $\chi^2$. 
The first one consists of neglecting the FFs uncertainty entirely, 
which leads to a $\chi^2/N_{\text{data}}$ distribution and a ${\cal P} (\alpha)$ 
peaked at values much larger than one. This is a clear indication for underestimated errors or 
and inadequate theory, and, in fact, the effective number of surviving replicas is $\sim 1$. 
For the other extreme, including the full 68\% C.L.\ uncertainties as estimated in \cite{dss17}, 
yields peaks below unity, characteristic of overestimated errors; another unwanted result
in applying the reweighting method.
The third option amounts to adjusting the C.L.\ such that ${\cal P} (\alpha)$
is peaked at unity as one would expect in a consistent reweighting of PDFs. 
The effective number of surviving replicas for this choice is found to be 310. 
In the following, we thus apply this criterion to include the uncertainties of FFs 
when reweighting PDFs based on SIDIS data.

%%%%%%%%%%%%%%%%
\section{Results of the reweighting exercise}
%%%%%%%%%%%%%%%%
%
In this section we present and discuss the outcome of the reweighting exercise 
based on kaon multiplicities in SIDIS in terms of the resulting set of PDFs 
and combinations of PDFs that quantify best the breaking of charge and flavor symmetry
we are interested in.
We also assess the impact of FFs in the reweighting process, and their
refinement through the iterative procedure outlined above.

\begin{figure}[t!]
\vspace*{-0.4cm}
\begin{center}
\hspace*{-0.4cm}
\epsfig{figure=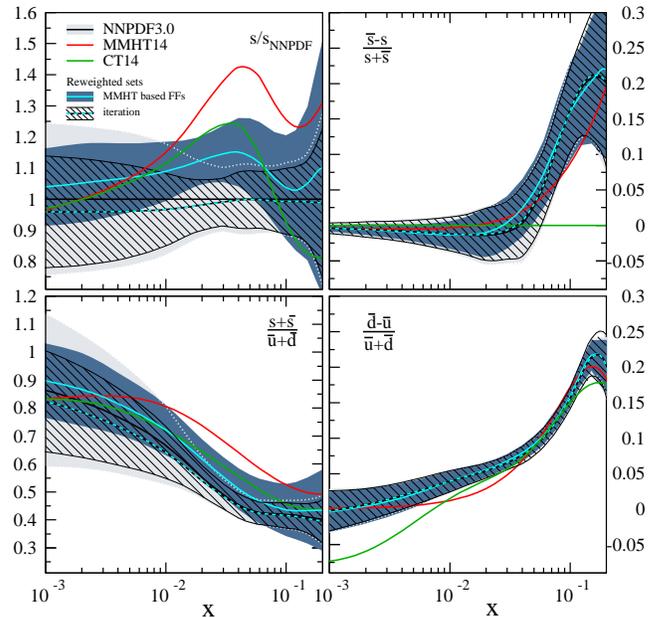,width=0.52\textwidth}
\end{center}
\vspace*{-0.5cm}
\caption{Reweighting of the strange quark distribution (upper left panel) 
and for the PDF combinations sensitive to charge (upper right panel) and 
flavor (lower panels) symmetry breaking using the DSS~17 set of kaon
FFs that is based on the MMHT~14 set of PDFs; see text.
The dashed light blue and black lines and the hatched areas represent 
the results of one iteration of the reweighting procedure 
and the corresponding uncertainty bands, respectively; see text.    
All results are shown at a scale of $Q^2=5\,\mathrm{GeV}^2$.
\label{fig:rew_MMHT}}
\end{figure}

Let us start with the strange quark density which is the main objective of
our reweighting exercise with kaon multiplicities in SIDIS.
In the upper left panel of Fig.~\ref{fig:rew_MMHT} we show different, recent
strange quark PDFs as a function of $x$ as obtained by 
MMHT~14 \cite{Harland-Lang:2014zoa} (red line), CT~14 \cite{Gao:2013xoa} (green line)
and NNPDF~3.0 \cite{Ball:2014uwa} (black line) to which all results are normalized.
Also included is the uncertainty band of the NNPDF~3.0 analysis (light grey band).
The first striking feature to notice in the plot are the differences between the 
current PDFs analyses for the strange quark distribution which can
exceed 40\% in case of MMHT~14 and even lie outside of the NNPDF3.0 uncertainty band 
in some region of $x$.

To demonstrate the impact of the {\sc Hermes} and {\sc Compass} kaon
multiplicities, the light blue curve and the blue shaded band
represent the result and the uncertainty estimate of a first
reweighting of the NNPDF~3.0 replicas for the strange quark PDF. 
This reweighting is performed by combining the replicas
with the DSS~17 set of kaon FFs to compute the weights $w_k$ by 
comparing to the SIDIS data. One should recall that the DSS~17 
global analysis makes use of the MMHT~14 set of PDFs and their
uncertainty estimates to determine the parton-to-kaon FFs.
We find that this exercise leads to a strange quark PDF that is
in between the starting set, i.e.\ NNPDF~3.0, and the MMHT~14 set
that is utilized by DSS~17. The corresponding uncertainty band
is significantly smaller than that provided by the original
NNPDF~3.0 analysis, indicating that a large number of the
NNPDF~3.0 replicas are excluded by the {\sc Hermes} and {\sc Compass} kaon
multiplicity data.

%
%%%%%%FIGURE 5
%

The above mentioned sizable differences between the strange quark densities
in recent PDF sets obviously questions the use of any particular PDF set
in the extraction of FFs, MMHT~14 in case of the recent DSS~17 global fit.
It even casts doubts on any result that is obtained by adopting these FFs.
In order quantify and even avoid this potential bias from the use of a specific 
set of PDFs inherent in the DSS~17 analysis of FFs, we performed
a second reweigthing exercise with the SIDIS kaon multiplicities.
To this end, we take the result from
the first reweighting as the input PDF set for a reanalysis of kaon FFs, following
the framework of DSS~17 in all other aspects. The so obtained set of FFs 
is then used for another round of reweighting of the NNPDF~3.0 replicas.
The outcome of this iterative procedure is the light blue and black dashed line
also shown in upper left panel of Fig.~\ref{fig:rew_MMHT}.
As can be seen, the iteration os now much closer to the  
original strange quark density of NNPDF~3.0 but, again, with smaller uncertainties
(hatched band). Notice that the error band resulting from one iteration 
of the reweighting procedure is different from the one without iteration of the FFs,
because in the comparison to data the improved set of FFs penalizes 
the NNPDF~3.0 replicas in the ensemble differently. 

The results of these exercises suggest that the {\sc Hermes} and {\sc Compass} kaon
multiplicity data appear to favor a PDF set more similar to NNPDF~3.0 than
to the other current sets, CT~14 and MMHT~14.
Indeed, the $\chi^2$ of the newly obtained FFs based on the
reweighted set of PDFs is significantly smaller, 1041.3 units,
than the original one, 1271.7, obtained in the DSS~17 fit \cite{dss17},
largely due to an improved description of the {\sc Hermes} and {\sc Compass}
data. We therefore conclude that the reweighting of the PDFs using 
SIDIS multiplicity data has a significant impact with regard to the strange quark distribution
and, in turn, the extraction of kaon FFs.
We also wish to stress that even though the first reweighting procedure
seems to be biased towards the MMHT~14 set of PDFs through the use of the DSS~17 FFs, 
the exercise helps to show that the method, in principle, allows for a significant departure 
from the mean value of ensemble of replicas one starts from.

The upper right panel of Fig.~\ref{fig:rew_MMHT} illustrates, 
in the same way as before, the  strange quark-antiquark or charge asymmetry,
which is rather small for all the different sets of PDFs
in most of the $x$ range, except at rather large values. 
As can be seen, in this case the result of the reweighting
is not as significant as for the strange quark PDF both in 
terms of a change in the asymmetry or in a reduction of the uncertainty band.
This clearly suggest a much weaker grip of the {\sc Hermes} and {\sc Compass}
data on this quantity beyond to what is 
already included in current global extractions of PDFs.

On the contrary, but perhaps not too surprisingly, a larger impact of the
reweighting is again found for the ratio between strange and 
non-strange light sea quarks which is shown in the 
lower left panel of Fig.~\ref{fig:rew_MMHT}.
The ratio measures the degree of SU(3) breaking in the proton and has been the
subject of considerable interest over the past decades.
Our result confirms a rather strong $x$ dependence in the strangeness suppression,
that weakens, i.e., the ratio approaches unity, towards smaller momentum fractions.
As for the strange quark distribution, the iterated reweighting procedure favors
the ratio obtained with the NNPDF~3.0 set of PDFs, but slightly 
modifies it and reduces the corresponding uncertainty.

As one anticipates, the reweigthing procedure using the
{\sc Hermes} and {\sc Compass} kaon multiplicity data has
very little impact on the difference between the $\bar{u}$ and $\bar{d}$ antiquarks 
as can be inferred form the lower right panel of Fig.~\ref{fig:rew_MMHT}.
SIDIS multiplicities for kaons are not competitive with the constraints 
on $\overline{u}$ and $\overline{d}$ already included in all the PDF fits.

%===========MS=========

%
%%%%%FIGURE 6
%
\begin{figure}[b!]
\vspace*{-0.4cm}
\begin{center}
\hspace*{-0.4cm}
\epsfig{figure=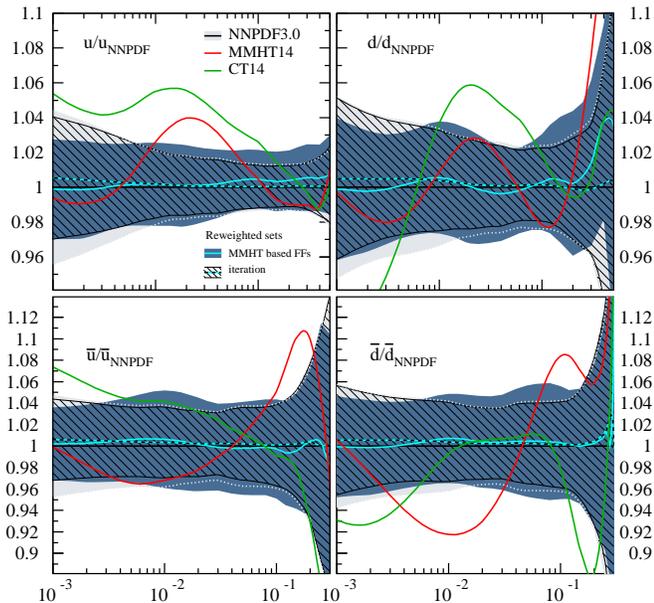,width=0.52\textwidth}
\end{center}
\vspace*{-0.5cm}
\caption{Same as Fig.~\ref{fig:rew_MMHT} but now for the non-strange light quark 
PDF flavors, again normalized to the corresponding mean value of the NNPDF~3.0 replicas.
\label{fig:rew2_MMHT}}
\end{figure}

For completeness, Fig.~\ref{fig:rew2_MMHT} presents the results of the 
reweighting exercises with and without iteration, performed in the
same way as discussed above, for the $u$, $d$, ${\overline u}$, 
and ${\overline d}$ quarks. For all these distributions, differences
among the different, current sets of PDFs are much smaller to begin with (at most
at a level of 10\%), i.e., the light quark flavors are 
already much better constrained than the strange quark PDF.
Even though the charged kaon multiplicities depend strongly on the
$u$ and ${\overline u}$ PDFs, and, in principle, the reweighting could 
have a non-negligible impact, it turned out that is has essentially none.

%%%%%FIGURE 7
%
\begin{figure}[t!]
\vspace*{-0.4cm}
\begin{center}
\hspace*{-0.4cm}
\epsfig{figure=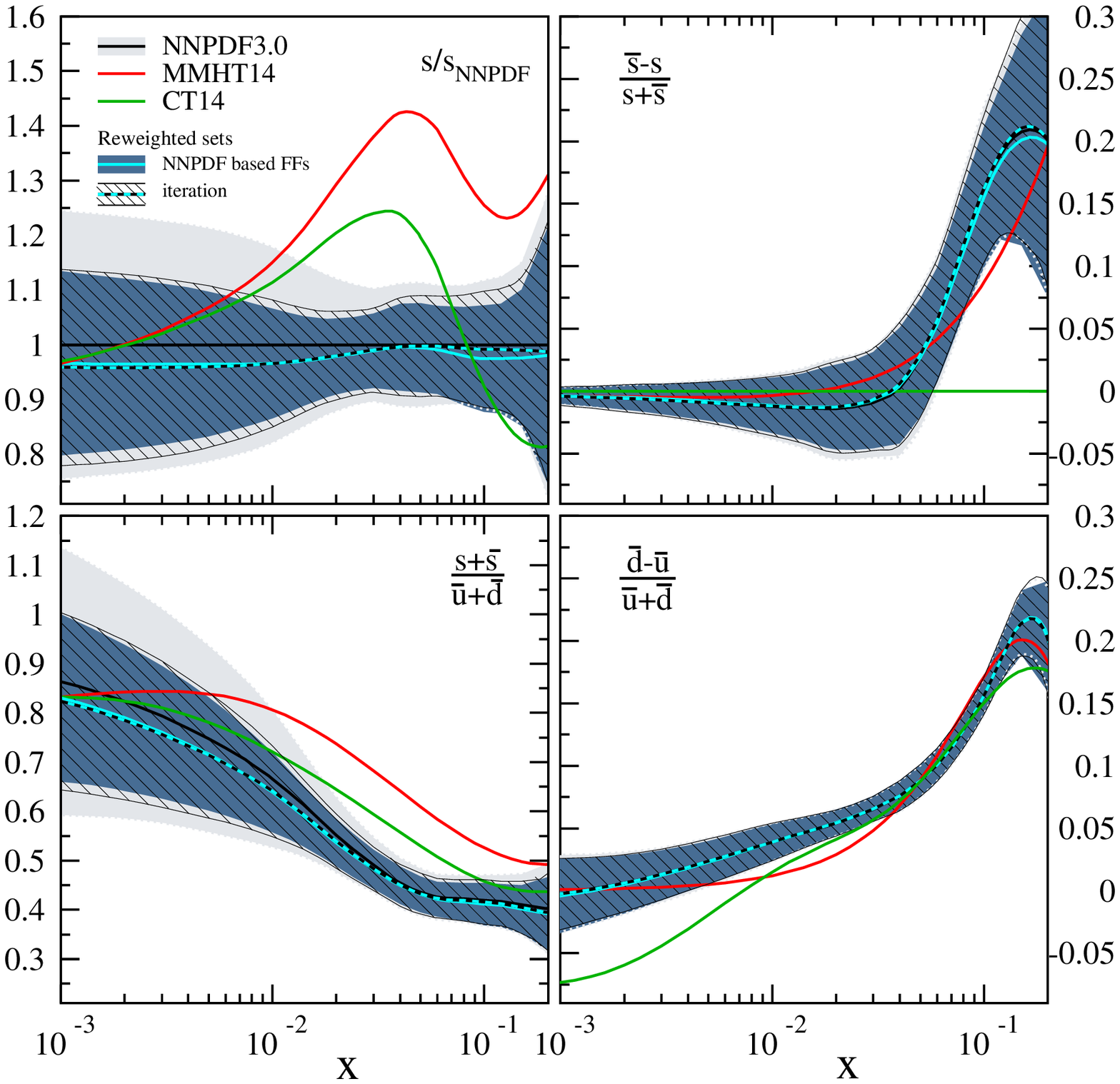,width=0.52\textwidth}
\end{center}
\vspace*{-0.5cm}
\caption{Same as Fig.~\ref{fig:rew_MMHT} but now starting from a variant of the DSS~17 
set of kaon FFs, extracted using the NNPDF~3.0 PDF set as input; see text. 
\label{fig:rew_NNPDF}}
%\end{figure} 
%
%%%%%FIGURE 8
%
%\begin{figure}[b!]
\vspace*{-0.4cm}
\begin{center}
\hspace*{-0.4cm}
\epsfig{figure=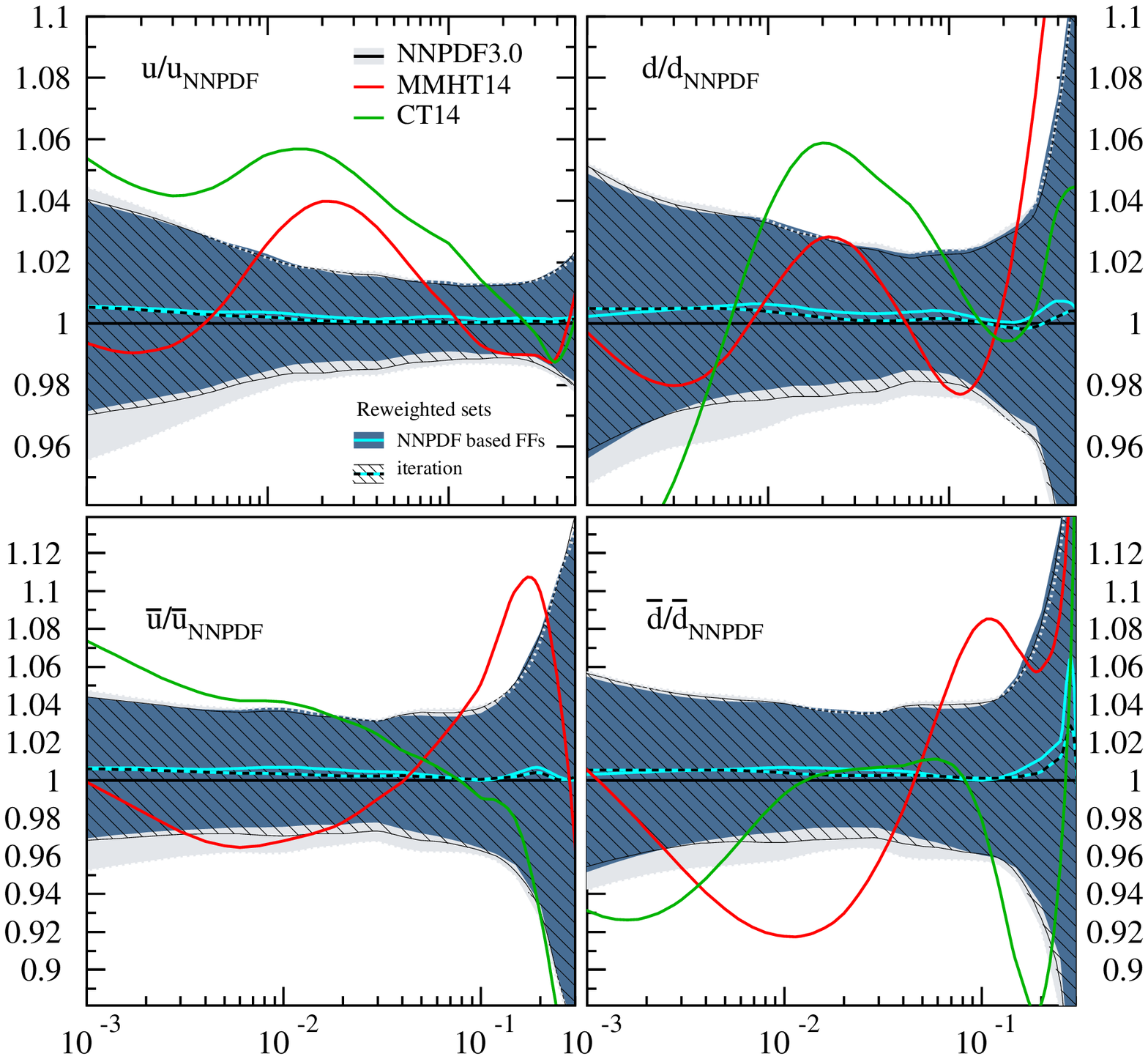,width=0.52\textwidth}
\end{center}
\vspace*{-0.5cm}
\caption{Same as Fig.~\ref{fig:rew2_MMHT} but now starting from a variant of the 
DSS~17 set of kaon FFs, extracted using NNPDF~3.0 PDF set as input; see text.
\label{fig:rew2_NNPDF}}
\end{figure} 

Finally, in Figs.~\ref{fig:rew_NNPDF} and \ref{fig:rew2_NNPDF} 
we repeat the reweighting exercises performed above 
but now with a set of kaon FFs that was obtained from a DSS~17-like global analysis 
using the NNPDF~3.0 set of PDFs as input.
In this case, the results of the first reweigthing (light blue curves) 
already resemble those obtained after the first iteration in Figs.~\ref{fig:rew_MMHT}
and \ref{fig:rew2_MMHT} above.
More importantly, the results now change only in a negligible way upon
iteration (light blue and black dashed line).
This highlights the robustness of the proposed iterative reweighting method 
and the rapid convergence to the same results despite starting from different
sets of FFs. The $\chi^2$ of the global fit of FFs in this approach, 1025.7 units, only
changes slightly from the one obtained above.

We note that recently, the simultaneous extraction 
of spin-dependent PDFs and FFs was studied in Ref.~\cite{Ethier:2017zbq} 
using an iterative Monte-Carlo method. The main objective of their analysis
was the helicity strangeness PDF. The precise data on unpolarized SIDIS multiplicities
and the crucial bias from unpolarized PDFs on the extraction of FFs 
are both not addressed in this paper.

%%%%%%%%%%%%%%%%
\section{Conclusions and outlook}
%%%%%%%%%%%%%%%%
%
SIDIS multiplicities depend crucially on parton distribution functions,
and, therefore, precise measurements of them should in principle constrain  
significantly PDFs in global analyses.
In practice, any use of SIDIS multiplicities also requires the
best possible knowledge of parton-to-hadron fragmentation functions,
whose determination in turn also involves PDFs.
Because of this complication, combined extractions of 
parton density and fragmentation functions have been suggested 
since the early days of the parton model.
The rather cumbersome technical details of such simultaneous 
global QCD fits have prevented so far their realization and implementation.

To circumvent these complications, we have proposed an
iterative strategy that is based on the reweighting approach, 
already routinely utilized for parton densities.
As a case study, we have implemented this procedure to analyze
the combined impact 
of recent charged kaon multiplicity data obtained
in deep-inelastic scattering 
on parton density and fragmentation functions.
It was demonstrated that the method converges very fast to a
stable result for new optimal sets of these non-perturbative
functions.

Our study has revealed a significant impact of the
kaon multiplicities on the strange quark distribution and its
uncertainties as well as on the ratio between strange and light
non-strange sea quarks in the proton. In the kinematic regime
relevant for the analyzed {\sc Hermes} and {\sc Compass} 
multiplicity data, we find that 
parton densities close to the NNPDF~3.0 set yield the optimum
description of the data in terms of $\chi^2$.

The recent availability of comparatively more precise 
charged pion multiplicities and parton-to-pion fragmentation functions 
\cite{deFlorian:2014xna}, and, eventually, extremely precise SIDIS data in 
a much wider kinematical range at a possible future Electron Ion Collider 
\cite{Accardi:2012qut}, highlights the increasing relevance of this kind of combined 
extractions of parton density and fragmentation functions, both of them pillars 
of the perturbative QCD description of hard processes based on the
notion of factorization.\\     

%%%%%%%%%%%%%%%%
\section*{Acknowledgments}
%%%%%%%%%%%%%%%%
%
We are grateful to Juan Rojo for his encouraging support and help
and to G.\ Schnell, E.C.\ Aschenauer ({\sc Hermes}) 
and F.\ Kunne, E.\ Seder ({\sc Compass}) 
for helpful discussions about their SIDIS measurements.
This work was supported in part by CONICET and ANPCyT.

%%%%%%%%%%%%%%%%%%%%%%%%%%%

%
\end{document}